# Infrared emissivity spectroscopy of a soda-lime silicate glass up to the melt

*Cristiane N. Santos*[1,*], *Domingos de Sousa Meneses*[1,2,§], *Valérie Montouillout*[1,2], *Patrick Echegut*[1,2]

[1] CEMHTI-CNRS, 1D Av. Recherche Scientifique, 45071 Orléans cedex 2, France

[2] Université d'Orléans, Faculté des Sciences, Avenue du Parc Floral, BP 6749, 45067 Orléans cedex 2, France

Corresponding authors, e-mail: *cristiane.n.santos@cnrs-orleans.fr; [§]desousa@cnrs-orleans.fr

**Abstract**

The short-range structure of an iron doped soda-lime glass was investigated by infrared emissivity spectroscopy from room temperature up to the melt. Quantitative information on the distribution of the $Q^n$ tetrahedral units was obtained by fitting the emissivity spectra using a dielectric function model (DFM). The DFM is based on causal Gaussian bands, associated with the stretching motions of the silicate tetrahedra. The changes in the absorption modes are related to the activation of a dynamical disorder that continuously increases with temperature. The obtained $Q^n$ speciation at room temperature is in good agreement with the magic-angle spinning nuclear magnetic resonance (MAS NMR) study. The distribution of the tetrahedral species undergoes slight changes with temperature, except during glass crystallization where $Q^4$ units increases, with a slight augmentation of $Q^2$ and $Q^4$ units in the melt. These results demonstrate the potentiality of infrared spectroscopy in the quantitative analysis of the polymerization degree of glasses and melts.

## 1. Introduction

Glasses are multifunctional materials that have found widespread use in industry, buildings, photonics and microelectronics. The soda-lime silica (NCS) system is the basic constituent of a wide range of commercially important glass products, yet there have been only few structural studies using vibrational spectroscopy.[1-3] To the knowledge of the authors, no information concerning the optical properties of the melt is available in literature for this system. Unveiling the structure and dynamical properties of such a glass melt should lead to a better understanding of the physical and chemical properties of the resulting glass, as well as of the radiative heat transfer occurring between the liquid and its surroundings during glass production. Optical in-situ high temperature studies offer the possibility for investigating these features from the glass to the liquid state, allowing to track the structural evolution during the melting process. From previous works, it is generally accepted that the short-range order in silicate glasses is composed of tetrahedral units with different number of bridging oxygens (BO), denoted by $Q^n$, where $n$ ($n$ = 0, 1, 2, 3 and 4) is the number of BO's. The relative amount of each unit in the glass network has been mostly determined using nuclear magnetic resonance (NMR) and Raman techniques.[4-7] Although infrared (IR) spectroscopy is sensitive to small-scale structural features, detailed IR spectra analysis have been neglected compared to Raman data, probably due to the complex nature of the IR spectrum, particularly the high-frequency part (800-1200 $cm^{-1}$) dominated by strong Si-O-Si and Si-O stretching vibrations. Due to this complexity, the majority of studies are limited to peak assignment without attempting to perform quantitative evaluation.[1,8] In the case of pure $SiO_2$ glasses, spectra decomposition and different interpretations of the vibration modes still generate some debate in the literature.[9-13] Compared to the glass structure, glass dynamics and melt properties are far less investigated, in part due to experimental challenges associated with high-temperature measurements. Among the few suitable techniques to characterize glass structure up to the melt, one can cite neutron and x-ray diffraction,[14-16] NMR,[4,5] infrared[17] and Raman[18-21] spectroscopies. Considering the vibrational techniques, in Raman spectra analysis the $Q^n$ speciation of glasses and melts requires a previous calibration using NMR data.[6,19,22] In spite of the good agreement found in this kind of analysis, a quantitative data analysis cannot be obtained without using calibration factors, computed by assuming a previously determined $Q^n$ distribution.[6,22,23] Up to now, the high-temperature IR studies only yielded qualitative information on glass dynamics, and the lack of information on IR absorption of molten silicates may preclude, for instance, the validation of *ab initio* calculations, as pointed out by Vuilleumier et al.[24] A recent study performed in our group allowed to follow quantitatively the structural dynamics of pure $SiO_2$ glass up to the melt.[25] In the present work, we demonstrate that quantitative information on the depolymerized silicate glass network and melt can be derived using IR emissivity spectroscopy, using a model NCS glass composition. The $Q^n$ speciation is achieved using the stretching region in the spectra. We take into account the dynamical disorder, as in the case of $SiO_2$ glass, and the results are compared with the predictions of the binary glass model. The obtained $Q^n$ distribution at room temperature is in good agreement with the NMR measurements and previous Raman and NMR studies.

## 2. Experimental Details

Glass samples with nominal composition $x$ $Fe_2O_3$ – 14.5 $Na_2O$ – 10.5 CaO - 75.0 $SiO_2$ (mol %), where $x$ = 0.005 or $x$ = 0.05, were prepared by Saint-Gobain Recherche, France. The glasses were melted at 1650°C during 2 h and then cooled down to few degrees above the glass transition temperature ($T_g$). The final measured compositions were 0.01 $Fe_2O_3$ – 13.8 $Na_2O$ – 10.7 CaO - 75.5 $SiO_2$ for the $x$ = 0.005 glass, and 0.04 $Fe_2O_3$ – 13.5 $Na_2O$ – 10.7 CaO - 75.7 $SiO_2$ for the $x$ = 0.05 glass, hereafter referred to as NCS glasses. The samples were cut into 7.5 mm-diameter disks with parallel faces, and optically polished down to a thickness of 1.986 mm measured by a mechanical comparator (Genève, Type 555) with a precision of ±2 µm. The spectra of other two glasses, amorphous $SiO_2$ (Spectrosil 2000, Saint-Gobain) and 13.2 $K_2O$ – 86.8 $SiO_2$ (mol%), hereafter respectively referred to as $a$-$SiO_2$ and KS13, were also measured for comparison with the NCS glass. The reflectivity and emissivity spectra were collected using a Bruker vertex 80v Fourier transform IR spectrometer, working under vacuum. In both configurations, the IR spectra are acquired at near normal incidence and therefore only transverse optical (TO) modes are excited. Data acquisition between 25-1600 $cm^{-1}$ was performed using liquid helium cooled bolometer (25-400 $cm^{-1}$) and DTGS (400-1600 $cm^{-1}$) detectors, with a resolution of 4 $cm^{-1}$ and accumulation of 64 scans. The emissivity spectra were acquired between 400-1600 $cm^{-1}$. In this spectral range, the silicate glass is opaque and its emissivity at room temperature is obtained indirectly using the Kirchhoff's law:

$$E = 1 - \rho \tag{1}$$

where $\rho$ is the glass reflectivity. For the high-temperature measurements, the emissivity was obtained by a direct method using an apparatus composed by a black body reference furnace, and a $CO_2$ laser (Diamond K500, Coherent Inc., 500 W) operating at 10.6 µm as heating source. To avoid axial thermal gradient, a beam-splitter allows to simultaneously perform back- and up-side sample heating. The sample compartment and the reference furnace are placed inside a chamber which is streamed with dry air. This guarantees a constant atmosphere during the measurement. The flux emitted by the reference, the sample and the surrounding are acquired and the sample emissivity spectra $E$ are obtained by the equation

$$E = \frac{FT(I_S - I_{RT})}{FT(I_S - I_{RT})} \frac{L(\omega, T_{BB}) - L(\omega, T_{RT})}{L(\omega, T_S) - L(\omega, T_{RT})} E_{BB}(\omega) \quad (2)$$

where the subscripts $BB$, $S$ an $RT$ stands for respectively the black body reference, the glass sample and the room temperature surrounding. $E_{BB}$ is the black body emissivity and thus its non-ideality is taken into account. $FT$ denotes Fourier transform, $I$ interferogram, $L$ the Planck's law and $\omega$ the wavenumber. Details on the experimental set up can be found in ref. 26. The sample temperature is calculated using equation (2) at the Christiansen point, where $E$ is nearly equal to one. The accuracy of this method was verified at the melting point of standard materials, such as $Al_2O_3$, giving an temperature uncertainty of ±20°C (not shown). The less iron doped NCS glass was characterized by $^{29}Si$ solid state magic-angle spinning (MAS) nuclear magnetic resonance (NMR). The spectrum was acquired at 59 MHz on a Bruker AVANCE 7.4 T (300 MHz) spectrometer equipped with a 4 mm double bearing MAS probehead spinning at 12 kHz. 128 scans were accumulated after a 45° pulse, using 3600s recycling delay. This delay ensures a complete relaxation of the magnetization, and then the quantitativity of the spectrum $^{29}Si$ chemical shifts were reported relative to TMS resonance. The spectrum was deconvoluted in individual gausso-lorentzian bands, using the Dmfit program.[27] Their (relative) integrated intensities were used to estimate the amount of the differently coordinated species.

## 3. Results

### 3.1 Thermal analysis

The thermogram of the glass sample is shown in Figure 1. Four major thermal events can be observed. The glass transition temperature ($T_g$) occurs around 835 K, the onset of glass crystallization ($T_{xi}$) at 965 K, and two crystallization peaks appears at 1066 K and 1180 K. After the crystallization process, named here as a characteristic temperature $T_{xf}$ around 1220 K, the first endothermal event can be related to the melt of an eutectic composition or phase transformation of the crystalline products. Based on the analysis of the phase diagram of the NCS system, Aboud et al.[28] have associated this event to the liquidus temperature of the glass, occurring around 1243 K. The onset of the glass melting ($T_m$) is found at 1480 K, and the melting peak ($T_{pm}$) occurs at 1643 K. According to previous crystallization studies of commercial 'float'[28,29] and NCS[30] glasses, for heat treatments between 1023-1273 K, the devitrification products are mainly $SiO_2$ crystalline polymorphs and devitrite ($Na_2O.3CaO.6SiO_2$), depending on treatment duration. Thus the crystallization peaks observed for the NCS glass can be associated with the formation of these phases. As we will discuss later, the crystallization process can be followed by in-situ emissivity measurements, as well as the structural changes in the short-range order associated with this process.

### 3.2 Infrared spectra analysis

The infrared emissivity spectra collected from above room temperature to 1566 K are shown in Figure 2. The room temperature emissivity (298 K) was obtained using the indirect method. According to equation (1) and Fresnel's law, the normal spectral emissivity is given by

$$E(\omega) = 1 - \left|\frac{\sqrt{\varepsilon(\omega)}-1}{\sqrt{\varepsilon(\omega)}+1}\right|^2 \qquad (3)$$

where ε is the dielectric function of the glass. The experimental spectra can be fitted on the basis of an adequate dielectric function model (DFM). De Sousa Meneses et al.[31,32] proposed a model consisting of causal Gaussian functions, taking into account the inhomogeneous broadening of the vibrational modes due to the structural disorder:

$$\varepsilon(\omega) = \varepsilon'(\omega) + \varepsilon''(\omega) = \varepsilon_\infty + \sum_j \left( g_j^{kkg}(\omega) + i\, g_j(\omega) \right) \qquad (4)$$

where $\varepsilon_\infty$ is the high-frequency dielectric constant that describes electronic contributions, $g_j$ is the causal Gaussian function and $g_j^{kkg}$ its Kramers-Kronig transform.[31,32] The fitting procedure allows to compute the dielectric function and therefore to derive all the optical functions of a material. This model was implemented using the software FOCUS,[33] where reflectivity and emissivity spectra are analyzed using the same dielectric function. The structure of various binary silicate[31,32] and borate[34] glasses have already been analyzed using this model. In silicate glasses, vibrations of the modifier cations in their oxygen coordination polyhedra are found in the 25-300 $cm^{-1}$ region, followed by the bending of the O-Si-O bonds (400-750 $cm^{-1}$) of the silicate network in the mid-infrared region. Here we will focus on the analysis of the high-frequency part of the spectrum, related to the stretching motions of the silicate network: Si-O-Si bond vibrations are found around 760-850 $cm^{-1}$, and Si-BO/Si-NBO bond vibrations occur between 850-1200 $cm^{-1}$.[1,31,35] For the room temperature emissivity of the NCS glass, a set of seven bands corresponding to the absorption due to these stretching vibrations was used to fit the experimental spectrum. The bands assignment was performed by comparing the spectrum of the NCS glass with those of *a*-$SiO_2$ and KS13 reference glasses, and with data from literature.[1,31,35] The infrared emissivity spectra of the NCS and reference glasses are shown in Figure 2a, together with the results from the fit. Examining the shape of the high-frequency region (800-1300 $cm^{-1}$), we can clearly see that the depolymerization degree changes for the NCS and KS13 glasses. The set of bands used to fit the spectrum of each depolymerized glass is presented in Figure 2b, based on the deconvolution proposed for the $a$-$SiO_2$,[25] also shown in Figure 2b. Depending on the glass composition, sets of six or seven bands were used. The present assignment constitutes a revised interpretation of the IR spectra of glasses. Previous studies did not take into account all the proper modes of $a$-$SiO_2$ and their temperature dependence.[8,31,35] The first evidence of the temperature behavior of the $a$-$SiO_2$ modes have been demonstrated recently in our group.[25] Noteworthy, a similar analysis was applied to the binary system $K_2O$ – $SiO_2$ and yielded a good description of the network depolymerization,[36] in agreement with the binary model and previous NMR studies. For the sake of clarity, the detailed band assignments are given in a subsequent section. The high-temperature analysis will focus on the high-frequency region.

**3.3 Infrared emissivity vs. Temperature**

The evolution of the emissivity spectra with temperature, acquired in the range 400 – 1600 $cm^{-1}$, is presented in Figure 3. The experimental data were fitted applying the same set of bands as for the room temperature spectral emissivity. Some of the fitting parameters and the imaginary part of the dielectric function ($\varepsilon''$), together with the Gaussian components used, are shown in Figure 4. As can be seen in Figure 3a, there is a decrease in intensity and a downwards shift in frequency with temperature. This trend changes at 824 K (near $T_g$) with a shift to higher frequency. In the bending region, the peak intensity varies slightly with temperature. Conversely, sudden changes occur in the entire spectral range above $T_g$, as shown in Figure 3b. The stretching peak position continues to shift to higher frequency at 954 K, and this trend is reversed at the next temperature (1077 K). Between 1184 – 1378

K, the upwards frequency shift is recovered. Moreover, both bending and stretching peak intensities increase, together with peak narrowing. These trends suggest that, in this temperature interval, the NCS glass undergoes a crystallization process, which is in agreement with the crystallization region $T_{xi}$ - $T_{xf}$ (965 - 1220 K) determined by DTA and IR measurements.[30] As indicated by the complexity of the thermal events (see Figure 1), more than one phase crystallizes and phase transformations take place upon heating. With increasing temperature, the formed crystallites melt and the general features are recovered, i.e. the bending and stretching peaks undergo an overall shift to lower frequency, broadening and the intensity diminishes. The temperature dependence of each stretching ($\nu_1$-$\nu_7$) vibration of the silicate network is depicted in Figure 5, where position, amplitude, full width at half maximum (FWHM) and the computed area of the modes are presented. Below $T_g$ no bond breaking takes place and consequently no significant changes in the relative quantities of $Q^n$ units should occur. Therefore, based on the temperature evolution of $a$-$SiO_2$[25], we assume that the changes in the stretching modes related to the $Q^n$ units are only due to the strengthening of a temperature-activated disorder. Above $T_g$, the effect of glass crystallization can be clearly seen in the amplitude and bandwidth of the component $\nu_4$, as well as in the shape of the stretching (800-1300 $cm^{-1}$) and bending (400 - 540 $cm^{-1}$) regions in the emissivity spectra (Figure 2a). To correctly fit the experimental data during this process, one additional band - $\nu_4'$ - was introduced in the spectra analysis at 1184 K and 1378 K. This band is related to stretching motions in a highly crystallized environment, presenting narrower bandwidth than the characteristic modes of the remaining glass. The kinetics of these phase transformations was not studied in this work. However, for each heating step the temperature was left to stabilize for at least 10-15 min, and no further peak evolutions were detected in this time scale. At higher temperatures (> $T_{xf}$), band broadening and intensity diminution of each component are first related to crystallite melting, and to bond loosening and breaking in the liquid state.

## 4. Discussion

### 4.1 Bands assignment and analysis

As mentioned before, the high-frequency room temperature spectra of the NCS and reference glasses were fitted using a set of five to seven absorption bands. One can observe that the deconvolution of the $SiO_2$ spectrum is already complex (Figure 3b), since five components are necessary to accurately reproduce the experimental data. Distinct and somehow controversial interpretations have been proposed, using a set of three, four or more bands.[12,13] A complete description of the structure dynamics of $a$-$SiO_2$ (from 4 K to 2500 K) is only achieved by using a set of five bands, which are related to the Si-O stretching motions.[25] The intertetrahedral stretching vibration from Si-O-Si bridges, with major contribution of the Si atoms, is located at 809 $cm^{-1}$. At higher frequency, the bands located at ca. 1059 $cm^{-1}$ and 1146 $cm^{-1}$ are associated with the stretching vibrations of Si-BO bonds in $Q^4$ tetrahedra. The latter band is related to Si-BO bonds experiencing dynamical disorder. This kind of disorder is induced by long range low frequency motions, related to the floppy modes of the silicate network.[37] Their population increases with temperature, leading to an enhancement of the dynamical disorder. Following this reasoning, the same set of absorption bands is used to represent the $Q^4$ units in our depolymerized glasses. It is generally accepted that a fully-polymerized tetrahedral silica network becomes gradually depolymerized upon addition of alkaline or alkaline earth oxide. If we consider the binary model, this depolymerization generates only $Q^3$ units up to an alkaline content of 1/3. Nevertheless, molecular dynamics (MD) simulations[38-41] and NMR data[42,43] shows that small quantities of $Q^2$ units are already present for lower modifier content. Moreover, it is known that higher field strength cations move the disproportionation reaction $2Q^3 \leftrightarrow Q^4 + Q^2$ to the right-hand side.[40] Therefore, the infrared spectra of the $a$-$SiO_2$ and KS13 glasses should sign the bands position of the $Q^4$ and $Q^3$ units, and in analogy with Raman[21,44] and IR[1] analysis, the band at lower frequency is a signature of $Q^2$ species. As can be seen in Figure 2a, the emissivity spectra of the glasses present distinct shapes, which corroborates with a dissimilar distribution of the tetrahedral units. Hence, in the NCS composition, a set of seven bands ($\nu_1$-$\nu_7$) was used to well reproduce the stretching region. The intertetrahedral stretching vibration from Si-O-Si bridges is found at 784 $cm^{-1}$ ($\nu_1$). The $\nu_2$ (930 $cm^{-1}$) and $\nu_3$ (972 $cm^{-1}$) modes are associated to stretching vibrations of Si-NBO bonds in $Q^2$ and $Q^3$ species, respectively. The stretching motions of Si-BO bonds in $Q^4$ units are assigned to the mode $\nu_4$(1029 $cm^{-1}$). Consistent with the analysis of $a$-$SiO_2$, the mode $\nu_6$ (1073 $cm^{-1}$) in the present glass can thus be associated to Si-BO contributions from both $Q^2$, $Q^3$ and $Q^4$ units in a dynamical environment. As we will elucidate hereafter, a transfer from vibrational modes occurring in a static environment to those experiencing dynamical disorder takes place with increasing temperature. These features account for the intensity changes observed in the IR stretching region. It is worth to note that in the analysis of Raman spectra, a band usually centered around 1050 $cm^{-1}$ has been attributed to Si-BO asymmetric stretching vibrations.[19,22,45] There is some controversy on its origin, which has also been assigned to stretching vibrations of $Q^3$ species.[46] Finally, the IR band at 1058 $cm^{-1}$ ($\nu_5$) is attributed to an external mode, related to intertetrahedral stretching vibrations, reflecting the state of the Si-O-Si bridges together with the mode $\nu_1$. In $a$-$SiO_2$, the area of these two modes decreases with increasing temperature up to the melt.[25] The low intensity mode $\nu_7$ (1218 $cm^{-1}$) is likewise of delocalized nature, and in $a$-$SiO_2$ glass it vanishes around 847 K.[25] This weak mode is also found in its crystalline analogue α-quartz around 1228 $cm^{-1}$.[8,13,35] As mentioned above, the deconvolution of Raman spectra and the use of calibration factors allow to determine the distribution of the $Q^n$ species in the glass structure.[19,22] Nevertheless, up to now few attempts have been made to obtain quantitative information on glass structure using infrared spectroscopy.[31,32,47] In the next section we demonstrate that quantitative information on the $Q^n$ distribution can also be extracted from the stretching modes in the infrared spectra. This analysis is performed using the present band assignment, based on the $a$-$SiO_2$ and its depolymerization with the addition of modifier cations, and in light of the binary model.

### 4.2 $Q^n$ speciation

Before going into details in the analysis of the vibration modes elucidated in the precedent section, some aspects of the silicate glass network will be discussed. As stated previously, the short-range order of the glass structure is described by the $Q^n$ units. A fully-polymerized tetrahedral silica network becomes gradually depolymerized upon addition of alkaline or alkaline earth oxide. If we consider that each alkaline/alkaline earth cation acts only as a modifier in the glass structure, no free oxygen is created and then for a binary system $x$ MO/$M_2O$ – (1- $x$) $SiO_2$, the number of oxygen atoms are calculated as follow: $n(NBO) = 2x$, $n(BO) = 2 - 3x$ and $n(O)=2-x$, where $n(O)$ is the total number of oxygen. For pure alkali binary glasses, it has been shown by IR, Raman and NMR studies that, depending on the alkaline ion, only $Q^4$ and $Q^3$ units are present for molar concentrations $x$ < 1/4.[14,36,42,43,45] Therefore according to the binary model (BM), the number of these units can be computed as: $n(Q^3) = 2x$, $n(Q^4) = 1 - 3x$ and $n(Q) = 1 - x$, the latter being the total number of $Q^n$ species. As already announced, the substitution of a part of alkaline cations by alkaline earth ones generally leads to the displacement of the disproportionation reaction $2Q^3 \leftrightarrow Q^4 + Q^2$ to the right-hand side.[40] This means that more $Q^4$ and $Q^2$ units will be present on an alkali-mixed glass. In this case, computing the theoretical values of the $Q^n$ units is less straightforward since the above BM equations are no longer strictly valid. On the other hand, for compositions where $x$ < 1/3, $Q^2$ units are present in small proportions.[7,14,42,43,45] The BM equations can thus be used to some extent assuming that Na and Ca are pure network modifiers.[42] The theoretical trends of the the $Q^n$ species – $t(Q^n)$ and dynamic mode $v_6$ – $t(v_6)$ were deduced from the BM and by applying the following rules: i) $Q^2$ units participate with two Si-NBO bonds to mode $v_2$ and two shared Si-BO bonds to the dynamic mode $v_6$; ii) $Q^3$ units with one Si-NBO bond to mode $v_3$ and 3/2 Si-BO bonds to mode $v_6$; iii) $Q^4$ units with $4(1-\delta)/2$ Si-BO bonds to mode $v_4$ and $2\delta$ Si-BO bonds to mode $v_6$. Here $\delta$ is defined as the fraction of BO's in $Q^4$ units under dynamical disorder and depends on temperature. Therefore, based on these statements, one has:

$$t(\nu_2) + t(\nu_3) = t(Q^2) + t(Q^3) = 2n(Q^2) + n(Q^3) = n(NBO) = 2x \quad \text{(5a)}$$

$$t(\nu_4) = t(Q^4) = 2(1-\delta)n(Q^4) \quad \text{(5b)}$$

$$t(\nu_6) = 2\delta n(Q_4) + \frac{3}{2}n(Q^3) + n(Q^2) \quad \text{(5c)}$$

$$t(\nu_4) + t(\nu_6) = 2(1-\delta)n(Q^4) + 2\delta n(Q_4) + \frac{3}{2}n(Q^3) + n(Q^2) = n(BO) = 2 - 3x \quad \text{(5d)}$$

$$t(v_2) + t(v_3) + t(v_4) + t(v_6) = n(NBO) + n(BO) = n(O) = 2 - x \quad (5e)$$

The BO's in $Q^2$ and $Q^3$ structural units are no longer constrained due to the presence of Si-NBO bonds, and they therefore experience dynamical disorder. It has been observed that in pure $a$-$SiO_2$, a nearly linear population increase of the mode associated to the dynamical disorder takes place with temperature.[25] In both $a$-$SiO_2$[49] and its crystalline counterpart $\beta$ cristobalite[49,50], the dynamical disorder has been attributed to out-of-phase rotations of different rings. We have proposed that in the infrared spectra, this disorder can be estimated by the ratio of the areas of the modes related to Si-O stretching vibrations in different environments[25]: (i) BO's in a dynamical local structure; (ii) BO's in a static local structure. At room temperature, dry and wet $a$-$SiO_2$ present values of $\delta$ near 0.15, and the dispersion around this value depends on the disorder and topology of the silicate network.[25] For the NCS glass, a value of $\delta\ (298\ K) = 0.18$ was used, in accordance with the results obtained for $a$-$SiO_2$. Above room temperature, the $\delta(T)$ values were computed by a linear fit of the ratio between the areas of the dynamic mode ($v_6$) and the $Q^4$ stretching vibrations ($v_4$). The obtained results will be discussed in the following section. Before computing the $Q^n$ distribution at room temperature, one should note that the area of the modes $v_2$, $v_3$ and $v_4$ are not directly related to the number of each $Q^n$ species. Indeed, to compare the above trends with the real $n_{exp}(Q^n)$ a few effects must be taken into account to extract these quantities. The first one is the infrared activity of the Si – BO and Si – NBO bonds. Tilocca and Leeuw[48] have shown by MD simulations that, for a $x$ = 0.15 NCS glass, the Si-NBO bond has a higher ionic character than the Si-BO bond. For the $Li_2Si_2O_5$ system, MD calculations resulted in values of 1.23 and 1.26 for the crystal and glass, respectively.[51,52]

In accordance with these results, here we define the ratio between the infrared activities of stretching motions of Si-NBO and Si-BO bonds as $\alpha_{BO}^{NBO} = 1.2$. To compare the theoretical trends described in the above equations (5a-5e) to the experimental results, the area of the modes were renormalized as follows:

$$r(v_2) = r(Q^2) = \frac{A(v_2)}{2a_{BO}^{NBO}C} = \frac{n_{\exp}(Q^2)}{C} \quad (6a)$$

$$r(v_3) = r(Q^3) = \frac{A(v_3)}{a_{BO}^{NBO}C} = \frac{n_{\exp}(Q^3)}{C} \quad (6b)$$

$$r(v_4) = r(Q^4) = \frac{A(v_4)}{2(1-\delta)C} = \frac{n_{\exp}(Q^4)}{C} \quad (6c)$$

$$r(v_6) = \frac{A(v_6)}{C} \quad (6d)$$

$$n_{\exp}(O) = r(v_2) + r(v_3) + r(v_4) + r(v_6) = 2 - x \quad (6e)$$

$$n_{\exp}(NBO) = 2r(v_2) + r(v_3) = 2x \quad (6f)$$

$$n_{\exp}(BO) = r(v_4) + r(v_6) = 2 - 3x \quad (6g)$$

where $C$ is a renormalization constant. Thus using $n_{exp}(Q^n)$ as described in relations 6a – 6c to compute the $Q^n$ speciation, we obtain at room temperature 4 ± 3%, 55 ± 3% and 41 ± 3% for $Q^2$, $Q^3$ and $Q^4$ units, respectively. This result has been confirmed by NMR measurements. The $^{29}$Si MAS NMR spectrum is shown in Figure 6 together with its deconvolution in individual Gaussian lines. According to their chemical shifts, respectively -81.1, -92.2 and -106.1 ppm, these three contributions can be attributed to $Q^2$, $Q^3$ and $Q^4$ species. The obtained speciation by IR and NMR are summarized in Table 1 and compared to other recent NMR, Raman, Neutron diffraction and MD calculations data on silicate glasses. As can be seen, there are some discrepancies between experiments, MD calculations[39] and reverse Monte-Carlo method (RMC).[16] This is probably due to the initial configuration or cooling-rate used to generate the glass structure in the MD calculations.[16,39] The changes in the structure and in the distribution of the $Q^n$ tetrahedra occurring with increasing temperature are discussed in the next section.

**4.4. High-temperature analysis**

As already introduced in section 3.3, the evolution of the emissivity spectra with temperature indicates remarkable changes in the intensity, bandwidth and position of the bending and stretching peaks. More information on the glass and melt structure can be accessed by analyzing each component separately. The amplitude of the modes related to stretching vibrations of the $Q^n$ units ($\nu_2$-$\nu_4$) and to the dynamical disorder ($\nu_6$) varies continuously with temperature, with sudden changes when glass crystallization takes place. Previous studies have shown that annealing NCS glasses between $T_{xi}$ – $T_{xf}$ leads to the formation of $SiO_2$ crystalline polymorphs and devitrite.[28-30] As $\nu_4$ is related to the $Q^4$ units, we evidence that the main crystalline phases formed at 1184 K are the $SiO_2$ polymorphs. The second crystallization peak ($T_{p2}$) in the DTA analysis (see Figure 1) occurs at the same temperature. This reflects a change in the short- and medium-range order of the glass structure, connected to the distribution of $Q^n$ units. These features can also be qualitatively seen in the bending peak at 450 cm$^{-1}$ (Figure 3b). In a more crystalline environment, the amplitude of the peak increases, together with peak narrowing, and approaches that of a more connected network such as the $a$-$SiO_2$. However, more information on the evolution of the bending vibrations would necessitate measurements below 400 cm$^{-1}$ for a complete description of these motions. Some information on the medium-range order can be obtained by analyzing modes $v_1$ and $v_5$. The structure of $a$-$SiO_2$ is composed of corner-sharing $SiO_4$ tetrahedra. The degrees of freedom of the structure allow the existence of internal modes, i.e. inside a single tetrahedron, and external modes, associated to the intertetrahedral Si-O-Si bridges. The area of these modes, attributed to the $v_1$ and $v_5$ bands, should decrease with increasing temperature due to bond loosening and breaking. Figure 7 shows the ratio between the areas of the modes $v_1$ and $v_5$ to the sum of the areas $A(v_1) + A(v_5)$, and the latter is presented in the inset of the same figure. The area ratios are nearly constant up to $T_g$. During glass crystallization, the spectral weight of $v_5$ increases as can be seen in the correlation of the area of the two modes. Moreover, the sum of their areas increases in this process, reflecting a more connected network. Except when crystallization takes place, these areas decrease with temperature, therefore attesting the external nature of the modes. Regarding the internal stretching vibrations, information can be extracted by examining the correlation between the total area of modes attributed to the $Q^n$ units – $\sum A(Q^n)$, and the area of the mode $v_6$, related to the dynamical disorder. As can be seen in Figure 8a, a nearly linear increase of $v_6$ is observed up to the melting of the glass, except during crystallization. The reverse behavior is found for $\sum A(Q^n)$. Moreover, no changes should occur in the relative quantity of $Q^n$ units up to $T_g$, evidenced by the sum $r(Q^2) + r(Q^3)$, which remains constant in this temperature interval (inset of Figure 8a). It thus becomes clear that, up to $T_g$, a linear fit of the ratio $A(v_4)/A(v_4) + A(v_6)$ gives a good approximation of the spectral weight transfer induced by the enhancement of the dynamical disorder. Defining $B$ as the obtained angular coefficient, we can hence compute $\delta(T) = 0.15 + BT$, where $\delta(298\,K) = 0.18$ as mentioned before. The linear fit of the ratio $A(\nu_6)/A(\nu_4) + A(\nu_6)$ is presented in Figure 8b, together with the obtained $\delta(T)$. The computed $Q^n$ speciation using equations *6a*-*6c* and $\delta(T)$ is shown in Figure 9. As expected, more $Q^4$ units are present during glass crystallization, together with an increase of $Q^2$ units, while the abundance of $Q^3$ diminishes, in accordance with the disproportionation reaction $2Q^3 \leftrightarrow Q^4 + Q^2$. Above $T_{xf}$, the formed crystallites start melting and the $Q^n$ distribution is close to that of the starting glass. These results show that the NCS glass network presents the same abundance of $Q^n$ species as the melt. Similar results were recently found for the NCS system using neutron diffraction. Melt speciation of alkali binary glasses with Raman spectroscopy shows the same trend, although some discrepancy is found in literature for similar alkali content.[6, 53] Using the above $Q^n$ distribution, we have computed the experimental ratio $NBO_{exp}/NO_{exp}$ employing equations 6e-6f. According to the composition, the theoretical ratio is given by $NBO_{teor}/NO_{teor} = 2x/2 - x = 0.28$. The obtained results are presented in Figure 10, together with the trend of the dynamic mode $t(\nu_6)$ and its experimental renormalized area $r(\nu_6)$, as predicted by equations 5c and 6d, respectively. The evolution of these quantities with temperature is in good agreement with the theoretical trends, and the ratio $NBO_{exp}/NO_{exp}$ is nearly constant up to the melt. The latter result indicates that no composition changes occurred during the measurements at high-temperature, a clear sign of the high sensitivity and potential of IR emissivity spectroscopy in the quantitative structural analysis of glasses and melts. Work on the kinetics of glass transformation and its impact on the short- and medium-range order over a wider spectral range (30-1600 $cm^{-1}$) are in progress by *in situ* crystallization of the NCS glass composition. Other silicate model glasses are under study and their structure dynamics can be derived, which demonstrates the general application of the present method.

## 5. Conclusions

We have shown that infrared emissivity spectroscopy is a powerful tool to analyze the glass dynamics and melt structure, and that the dynamical disorder plays a crucial role on the spectral weight of the stretching components with increasing temperature. Taking this dynamic aspect into consideration, we could demonstrate that quantitative information on glass polymerization degree and consequent speciation of the $Q^n$ tetrahedral units up to the melt can be derived using IR spectroscopy. The obtained results are in good agreement with the NMR analysis of the same glass composition, and with NMR, Raman and MD studies on binary and mixed-alkali silicate glasses. The predictions of the binary model are still a good approximation for the studied composition.

## Acknowledgments

The authors are thankful to S. Ory and G. Matzen for the DSC measurements. This work was supported by the French national research agency (ANR) under the program MatePro2008, Postre project (08-MAPR-0007).

## List of Figures

Fig. 1 – Differential scanning calorimetry (DSC) thermogram of the NCS glass with a heating rate of 10°C/min.

Fig. 2 – (a) Room temperature emissivity spectra of the NCS and model glasses ($a$-$SiO_2$ and KS13). Symbols are experimental data, and lines the obtained fit. (b) Set of absorption bands related to the stretching vibrations of the silicate network.

Fig. 3 – Normal directional emissivity spectra of the NCS glass as a function of temperature.

Fig. 4 – (a) Experimental emissivity spectra (symbols) and fit from the model (lines) at 298 K, 1184 K and 1566 K. (b) Imaginary part of the dielectric function and deconvoluted Gaussian bands.

Fig 5. Gaussian bands parameters as a function of temperature: (a) position, (b) amplitude, (c) full width at half maximum (FWHM) and (d) computed area.

Fig 6 – $^{29}$Si MAS NMR spectrum of the NCS glass, and Gaussian line shapes for each $Q^n$ unit.

FIG 7 – Temperature dependence of the area ratios of the modes $v_1$ and $v_5$. The inset shows the sum of the areas.

Fig. 8 (a) Temperature dependence of the ratios between the total area of the modes attributed to the $Q^n$ units -$\sum A(Q^n)$ - and the area of the dynamic mode $v_6$ - $A(v_6)$. The inset presents the sum of the renormalized areas of the modes attributed to the $Q^2$ and $Q^3$units. (b) Linear fit of the areas ratio $A(v_4)/A(v_4)+A(v_6)$, together with the computed $\delta(T)$ values. Lines are guides to the eyes.

Fig. 9 – Temperature dependence of the $Q^n$ distribution in the NCS.

Fig. 10 – Temperature dependence of the experimental ratio $NBO_{exp}/NO_{exp}$, along with the theoretical prediction and renormalized area of the dynamic mode $v_6$. The line is a guide to the eyes.

List of Tables

Table 1: $Q^n$ speciation in NCS and silicate glasses

## Tables

Table 1

| Glass composition (mol%) | $Q^2$ (%) | $Q^3$ (%) | $Q^4$ (%) | Method | Reference |
|---|---|---|---|---|---|
| 0.04 $Fe_2O_3$ – 13.5 $Na_2O$ – 10.7 CaO - 75.7 $SiO_2$ | 4 ± 3 | 55 ± 3 | 41 ± 3 | IR | This work |
| 0.01 $Fe_2O_3$ – 13.8 $Na_2O$ – 10.7 CaO - 75.5 $SiO_2$ | 3.4 ± 3 | 58.0 ± 2 | 38.0 ± 2 | NMR | This work |
| 15.5 $Na_2O$ – 9.9 CaO - 74.6 $SiO_2$ | 2.7 ± 1.0 | 62.7 ± 1.0 | 34.6 ± 1.0 | NMR | 7 |
| 23.8 $Na_2O$ - 4.7 CaO - 71.5 $SiO_2$ | 0 | 73 ± 5 | 27 ± 5 | NMR | 42 |
| 10 $Na_2O$ - 15 CaO - 75 $SiO_2$ | 9.9 | 44.4 | 44.8 | MD | 39 |
| 11.78 $Na_2O$ - 14.86 CaO - 73.36 $SiO_2$ | 10.17 | 41.75 | 46.7 | Neutron+RMC | 16 |
| 25 $Na_2O$ - 75 $SiO_2$ | 2 | 48 | 50 | Raman | 6 |

## Figure Captions

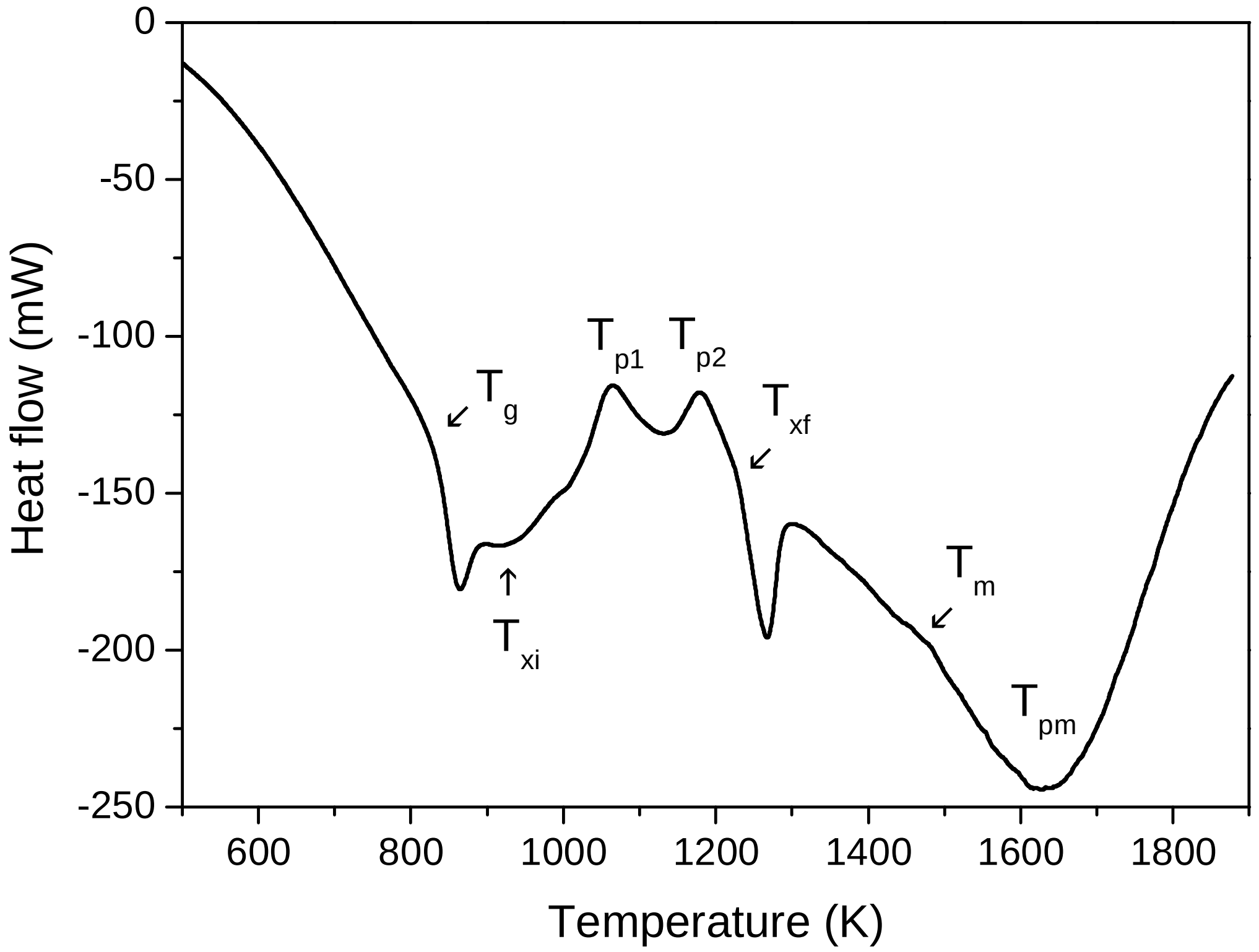

Fig. 1 (Santos et al.)

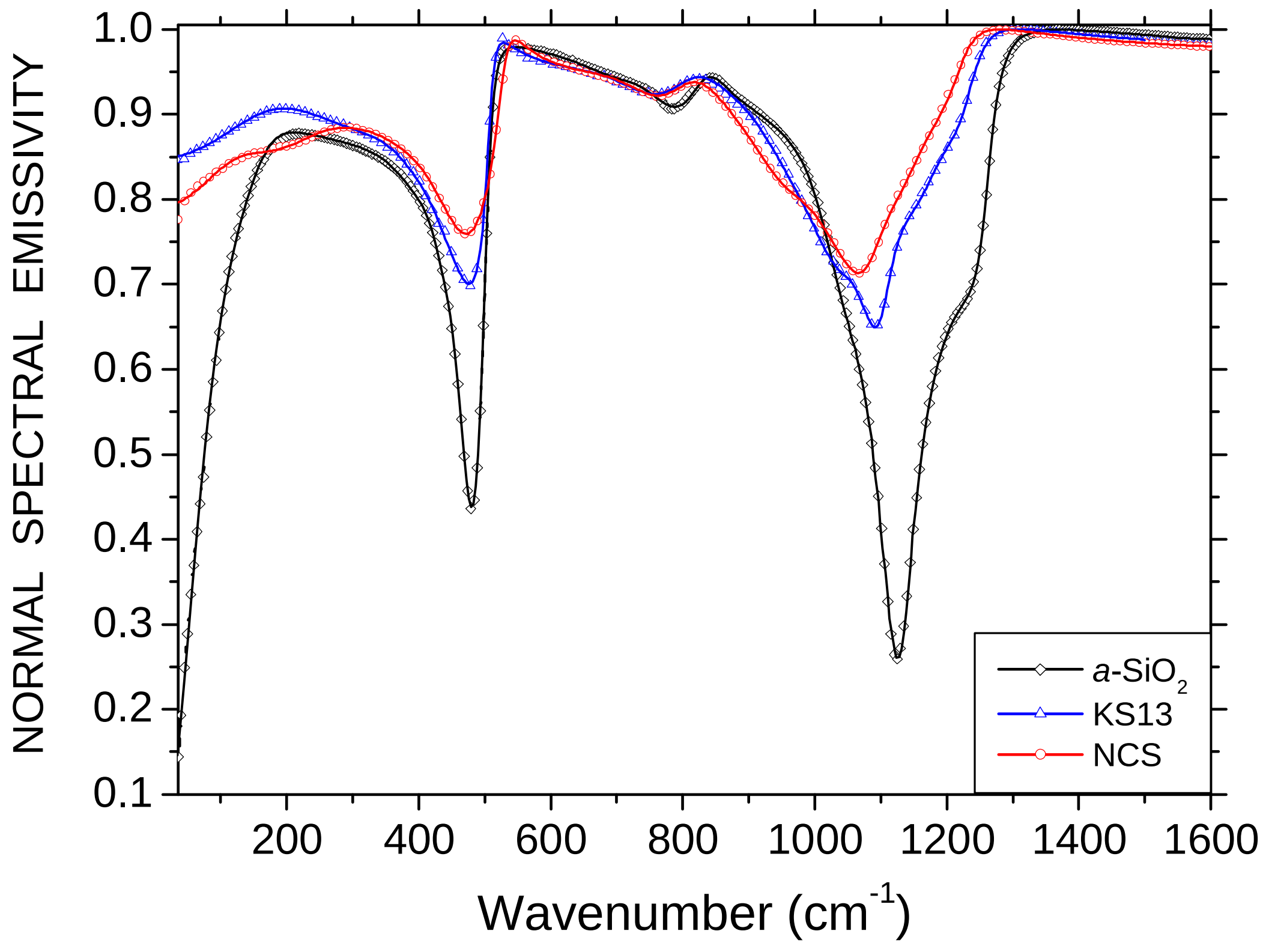

Fig. 2a (Santos et al.)

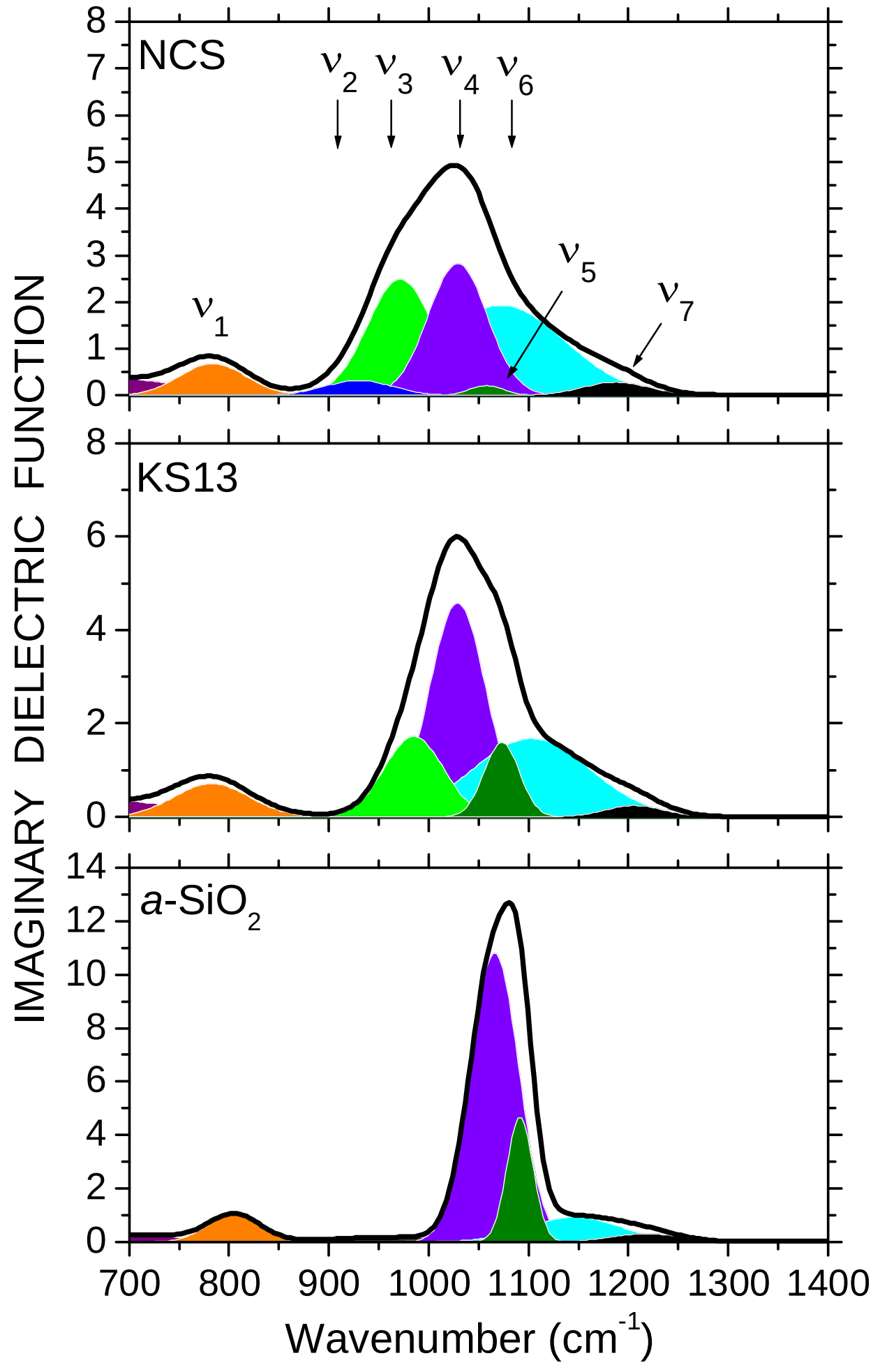

Fig. 2b (Santos et al.)

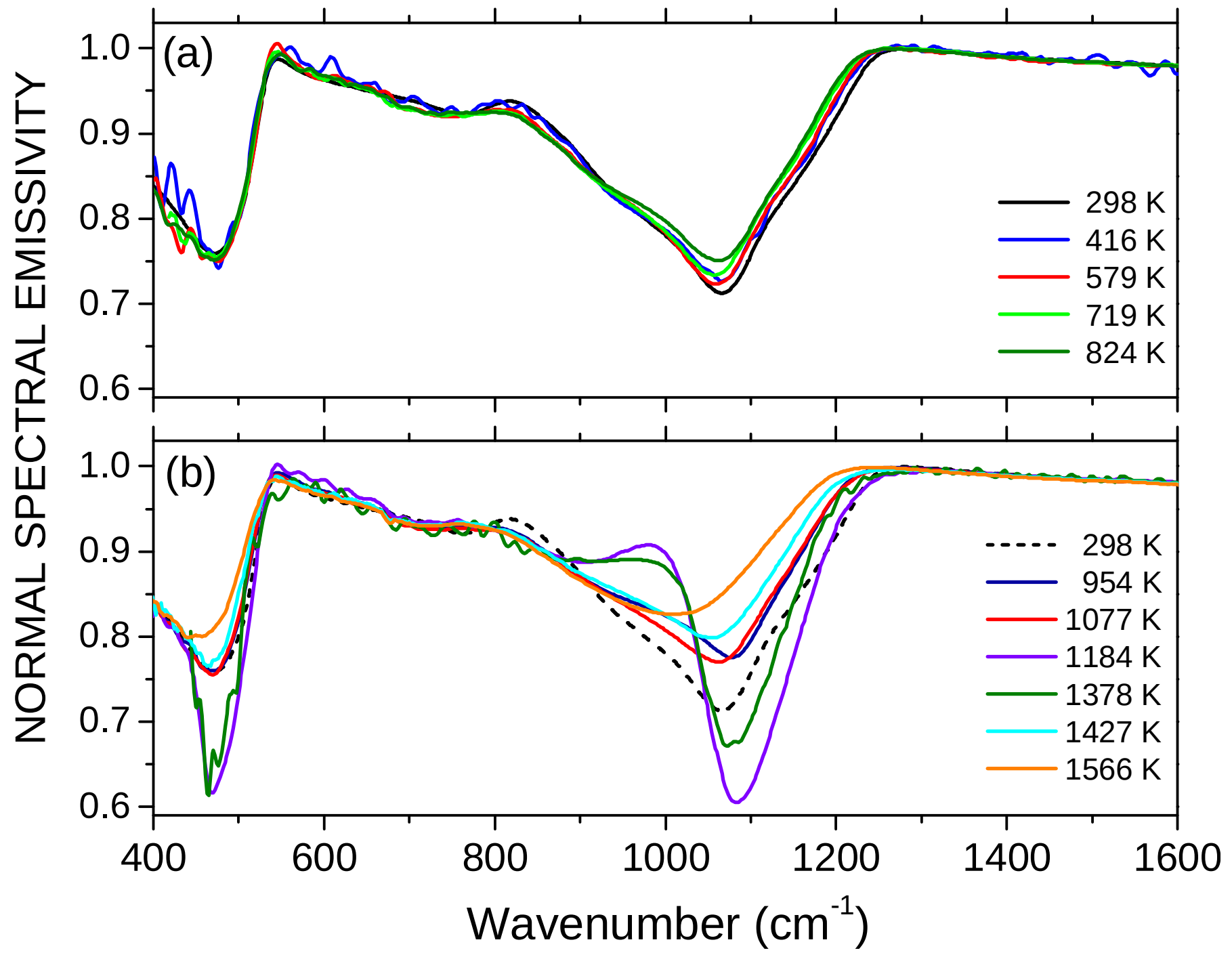

Fig. 3 (Santos et al.)

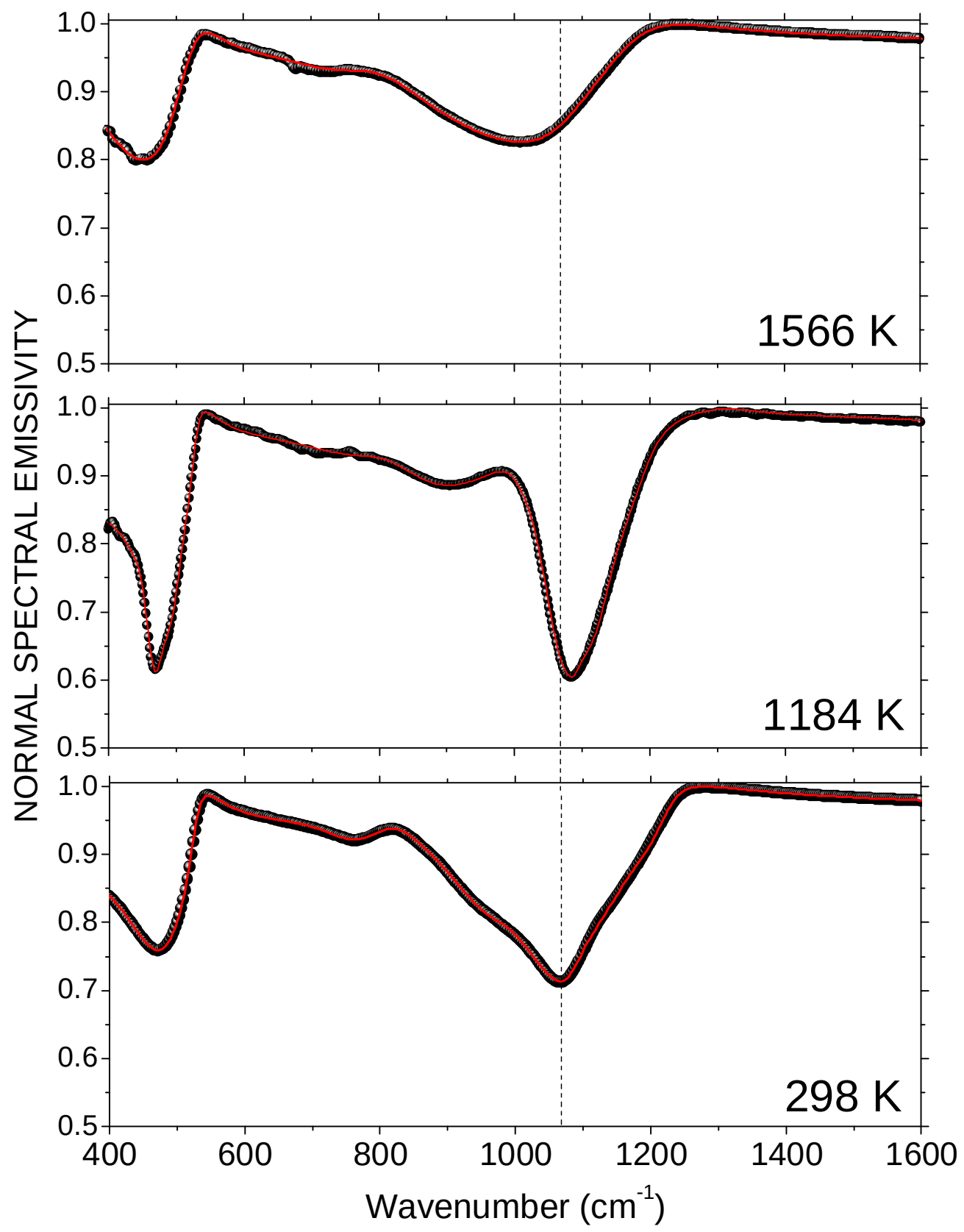

Fig. 4a (Santos et al.)

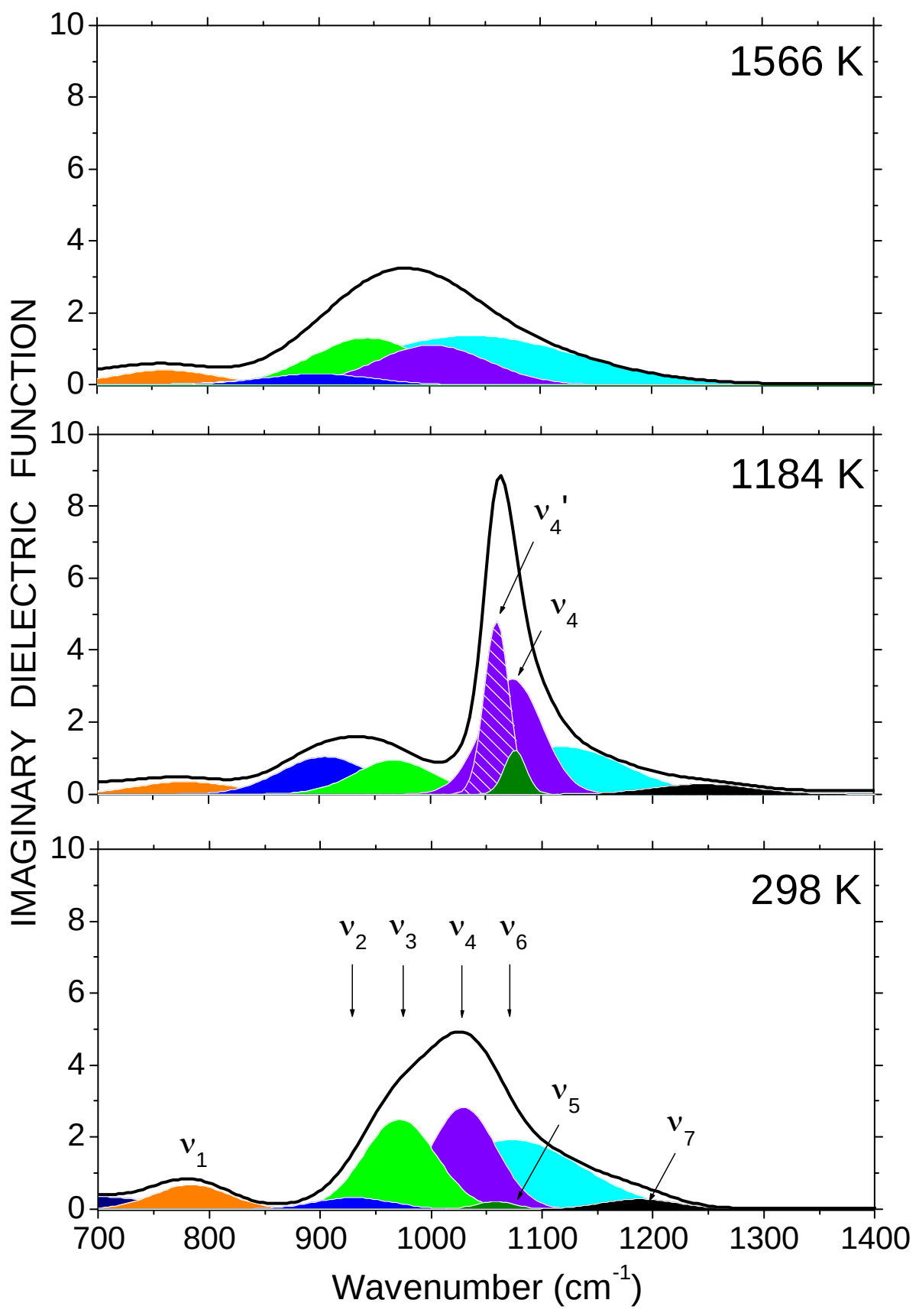

Fig. 4b (Santos et al.)

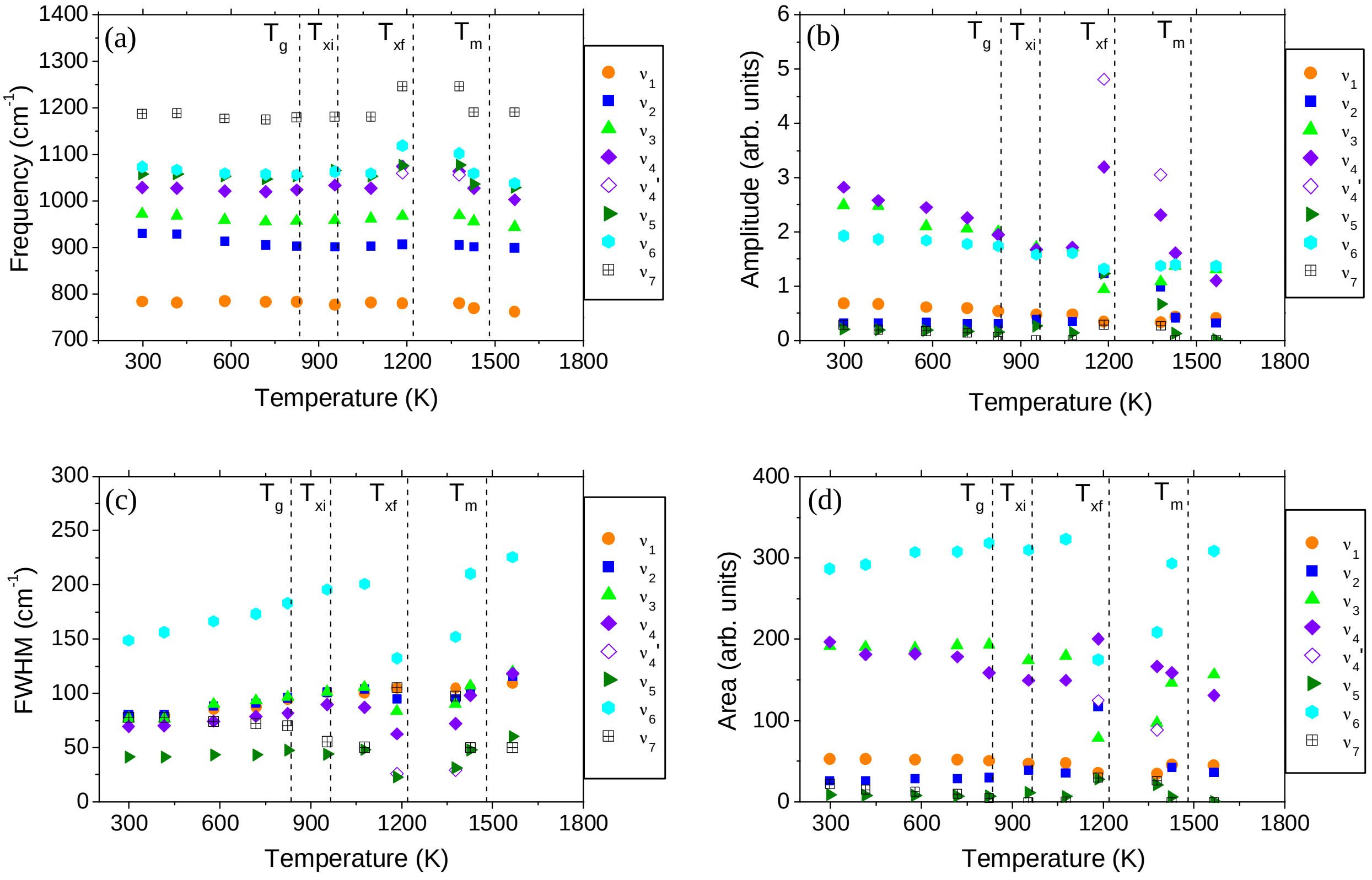

Fig. 5 (Santos et al.)

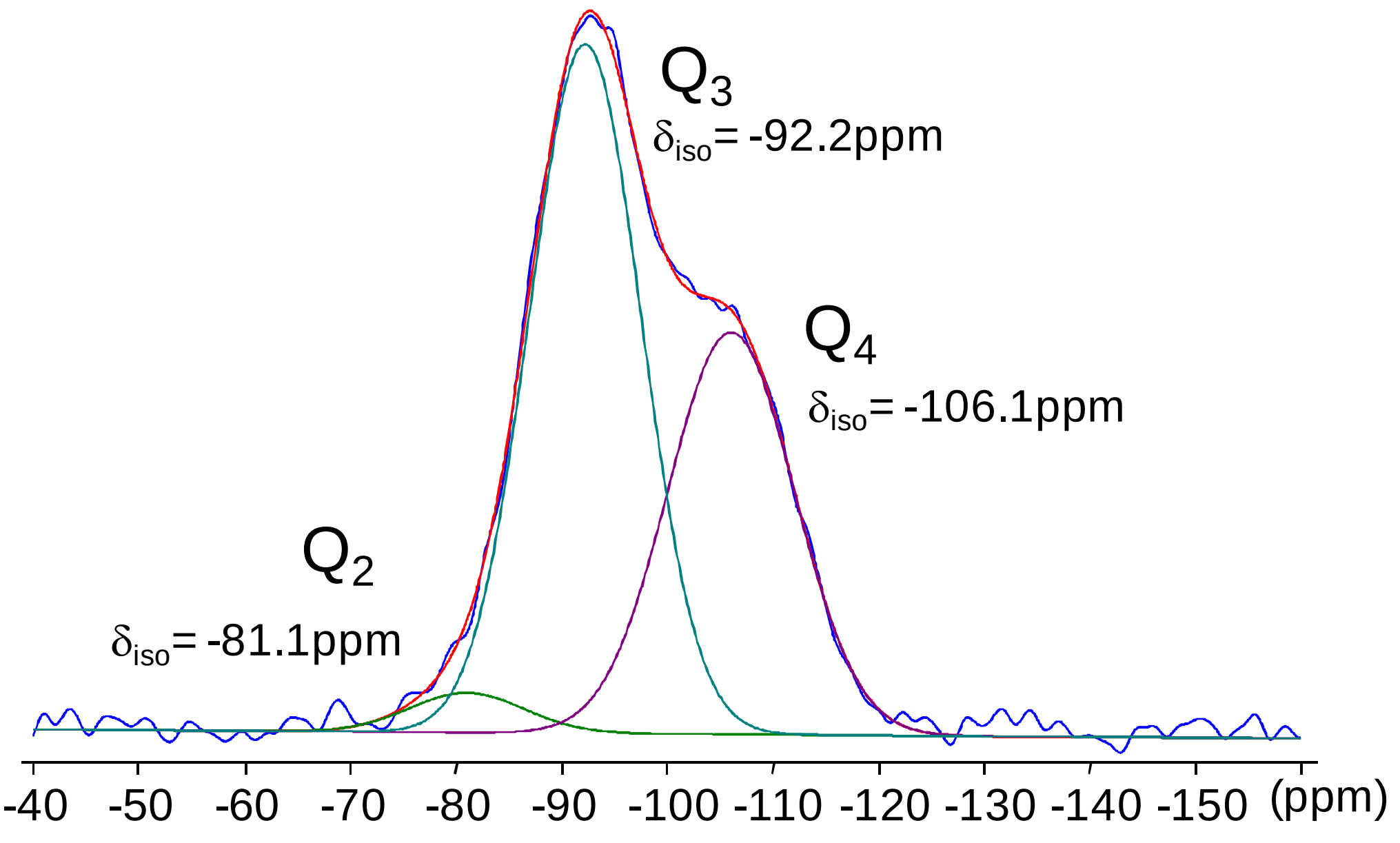

Fig. 6 (Santos et al.)

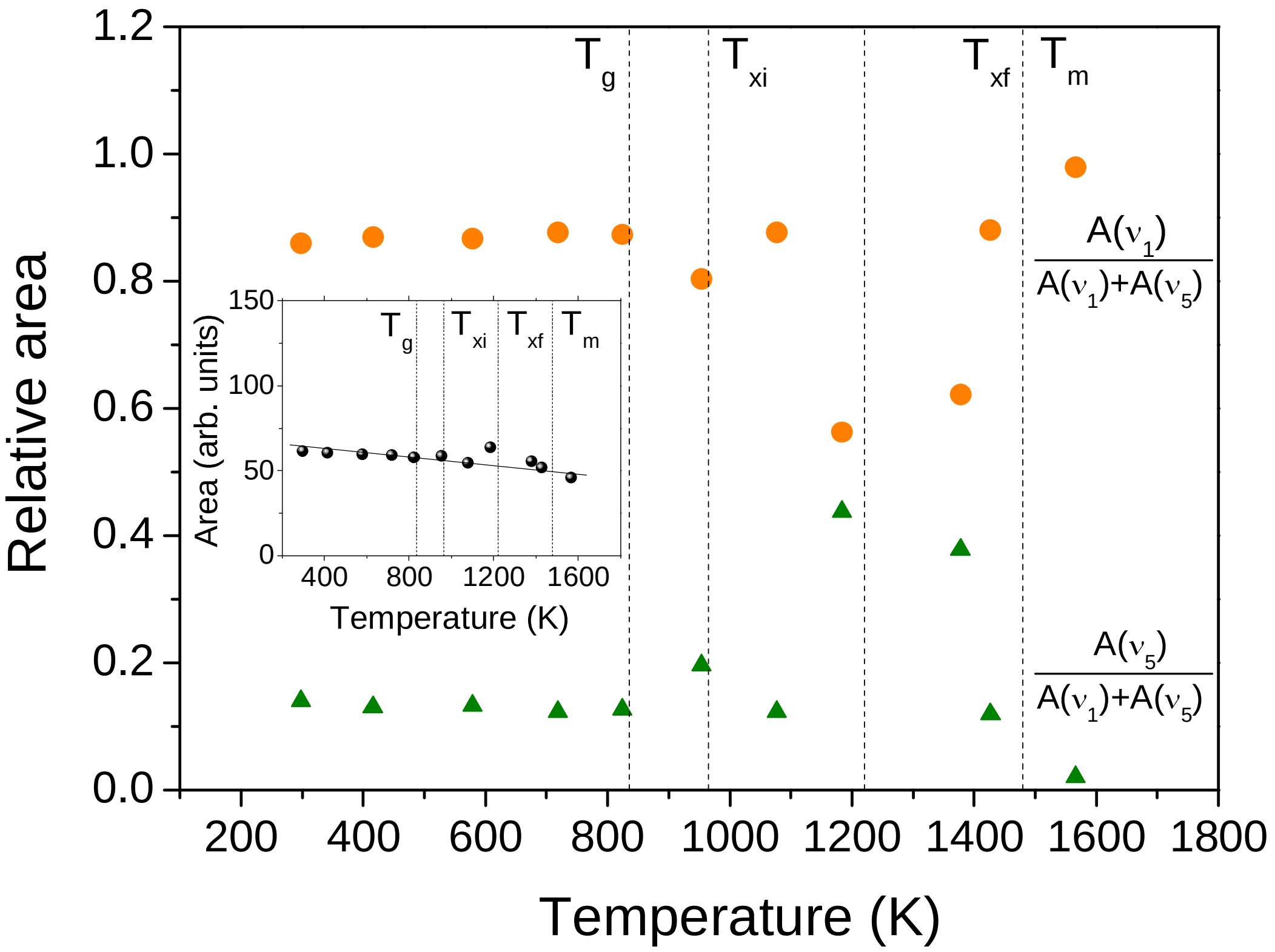

Fig. 7 (Santos et al.)

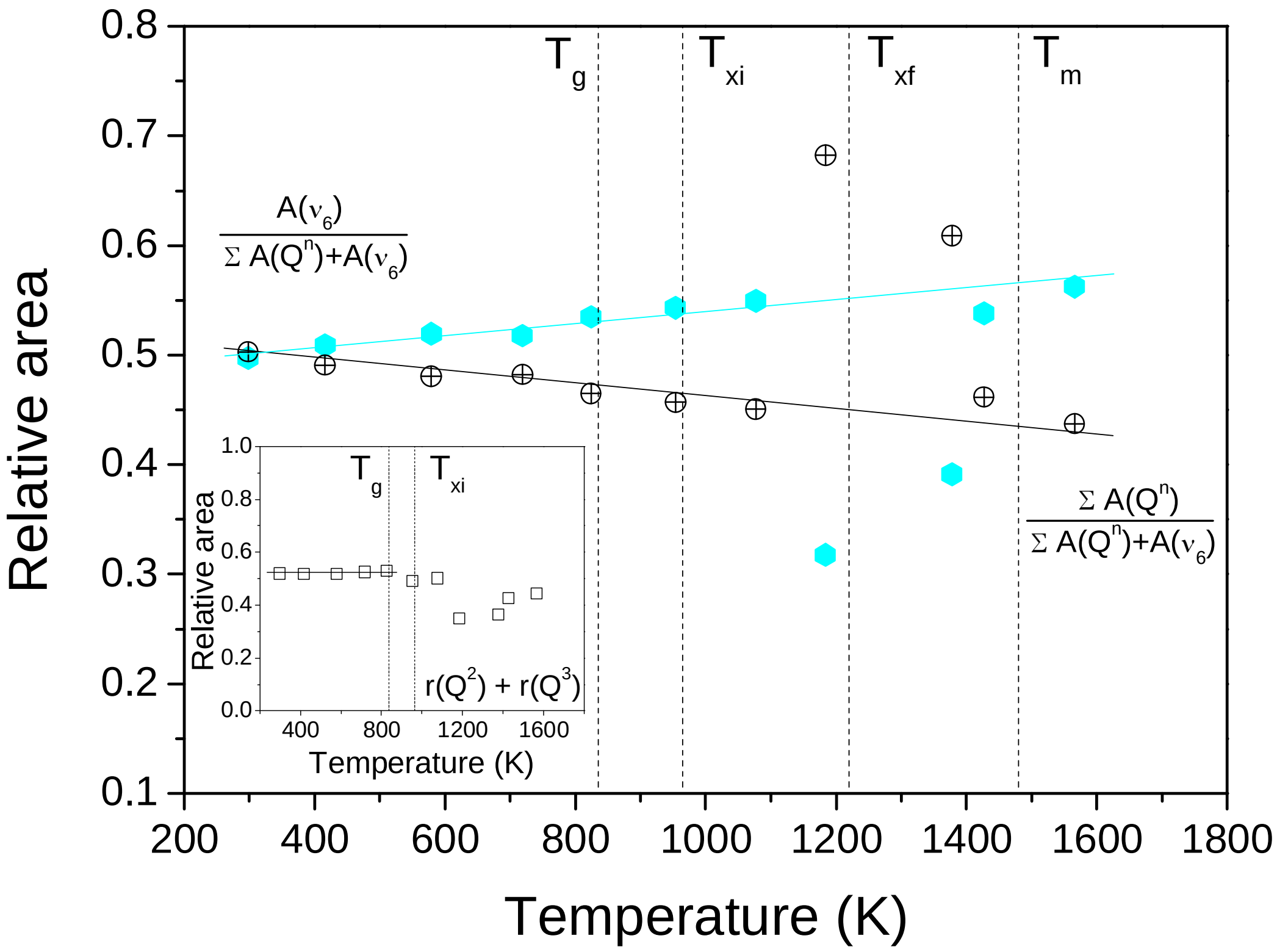

Fig. 8a (Santos et al.)

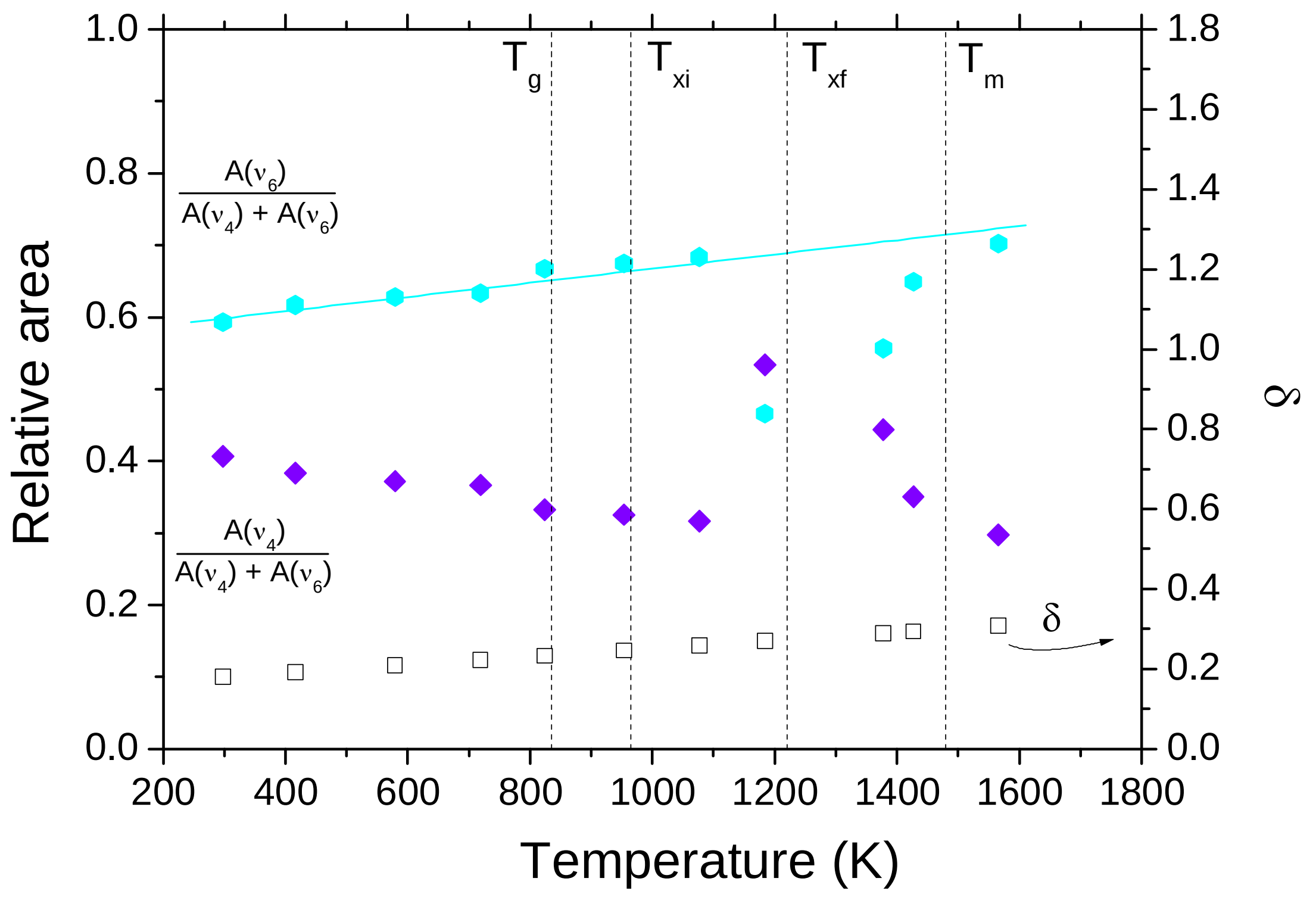

Fig. 8b (Santos et al.)

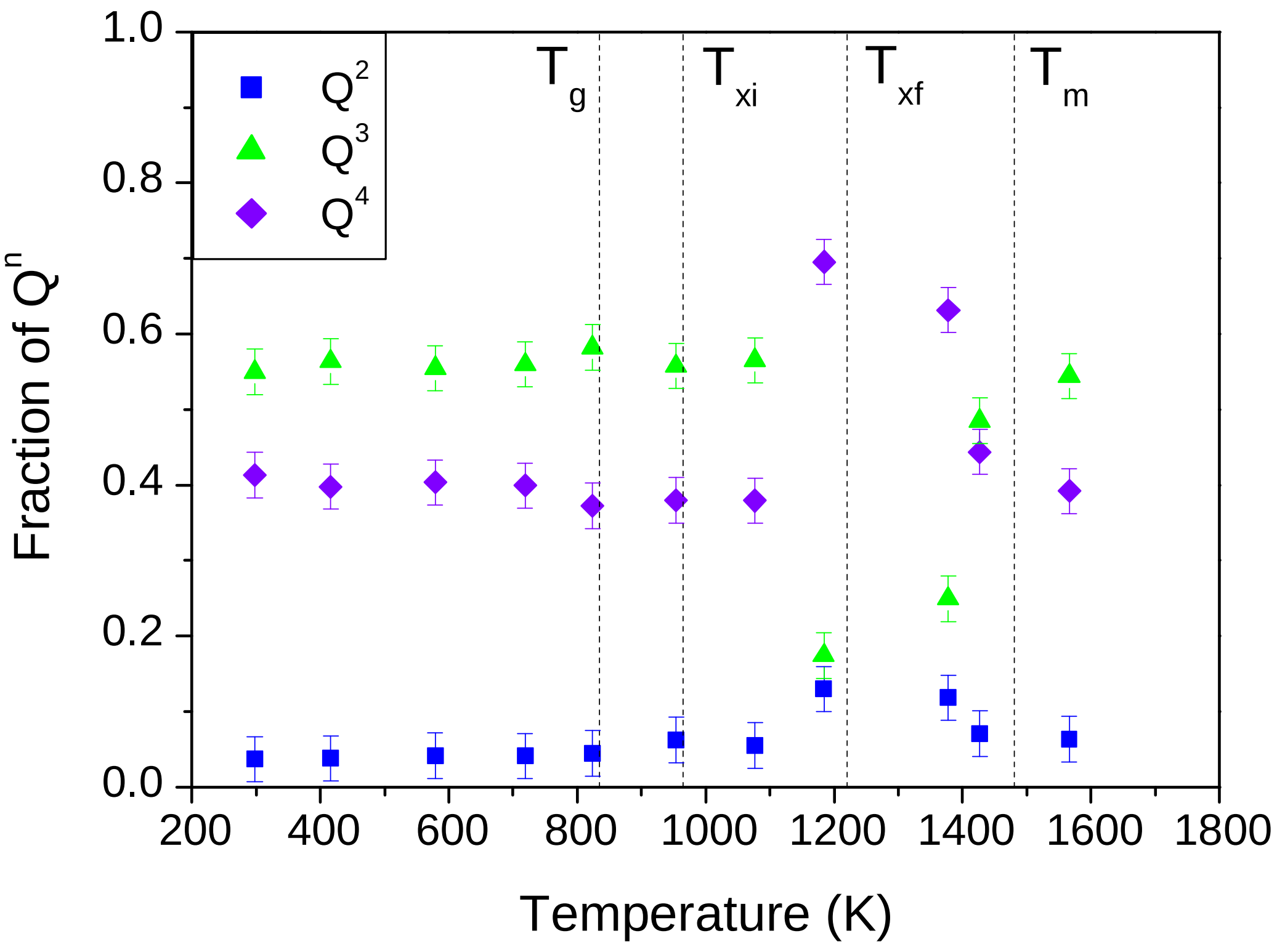

Fig. 9 (Santos et al.)

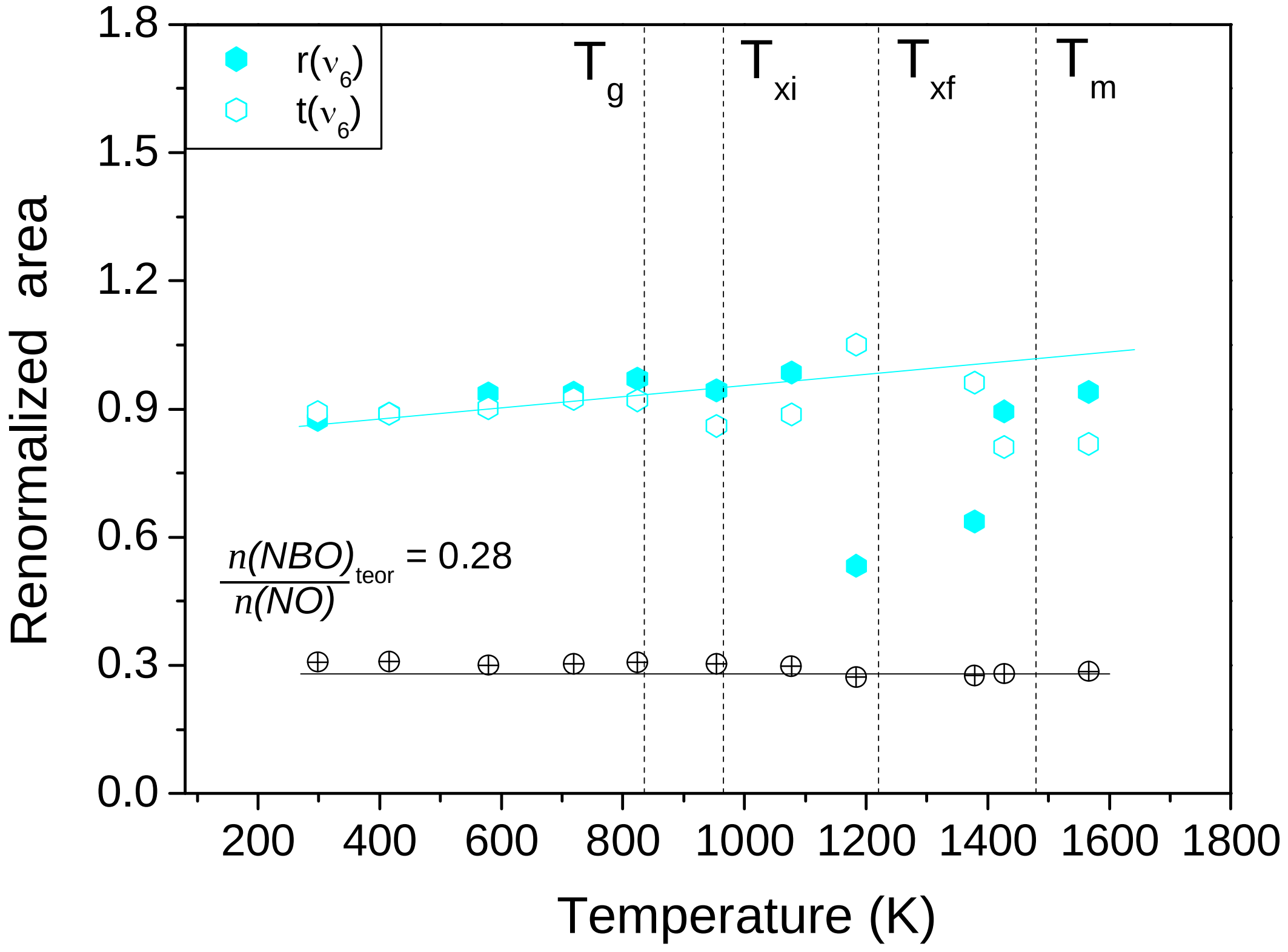

Fig. 10 (Santos et al.)